\documentclass[reprint, prl, superscriptaddress, amsmath, amssymb, floatfix]{revtex4-2}

\usepackage{graphicx, dcolumn, bm, svg, upgreek, xfrac, float, mathtools}
\usepackage[normalem]{ulem}
\usepackage[colorlinks=true, allcolors=blue]{hyperref}
\usepackage{bookmark}

\newcommand{\bra}[1]{\left< #1 \right|}
\newcommand{\ket}[1]{\left| #1 \right>}

\begin{document}

% Title
\title{Rabi oscillations of a monolayer quantum emitter driven through its excited state}
\date{\today}

% Author List
\author{Victor~N.~Mitryakhin}
    \affiliation{Institute of Physics, Carl von Ossietzky University of Oldenburg, 26129 Oldenburg, Germany}
\author{Ivan~A.~Solovev}
    \affiliation{Institute of Physics, Carl von Ossietzky University of Oldenburg, 26129 Oldenburg, Germany}
\author{Alexander~Steinhoff}
    \affiliation{Institute of Physics, Carl von Ossietzky University of Oldenburg, 26129 Oldenburg, Germany}
\author{Jaewon~Lee}
    \affiliation{Department of Physics, Stockholm University, 10691 Stockholm, Sweden}
\author{Martin~Esmann}
    \affiliation{Institute of Physics, Carl von Ossietzky University of Oldenburg, 26129 Oldenburg, Germany}
\author{Ana~Predojevi\'{c}}
    \affiliation{Institute of Physics, Carl von Ossietzky University of Oldenburg, 26129 Oldenburg, Germany}
    \affiliation{Department of Physics, Stockholm University, 10691 Stockholm, Sweden}
\author{Christopher~Gies}
    \affiliation{Institute of Physics, Carl von Ossietzky University of Oldenburg, 26129 Oldenburg, Germany}
\author{Christian~Schneider}
    \affiliation{Institute of Physics, Carl von Ossietzky University of Oldenburg, 26129 Oldenburg, Germany}
    
% Abstract
\begin{abstract}

The interaction of a quantum two-level system with a resonant driving field results in the emergence of Rabi oscillations, which are the hallmark of a controlled manipulation of a quantum state on the Bloch sphere. This all-optical coherent control of solid-state two-level systems is crucial for quantum applications. In this work we study Rabi oscillations emerging in a  WSe$_2$ monolayer-based quantum dot. The emitter is driven coherently using picosecond laser pulses to a higher-energy state, while photoluminescence is probed from the ground state. The theoretical treatment based on a three-level exciton model reveals the population transfer between the exciton ground and excited states coupled by Coulomb interaction. Our calculations demonstrate that the resulting exciton ground state population can be  controlled by varying driving pulse area and detuning which is evidenced by the experimental data. Our results pave the way towards the coherent control of quantum emitters in atomically thin semiconductors, a crucial ingredient for monolayer-based high-performance, on-demand  single photon sources. 
\end{abstract}

\maketitle

%\section{Introduction}

Solid-state single-photon emitters are one of the fundamental building blocks of integrated quantum technology applications  \cite{waks_quantum_2002,kimble_quantum_2008, obrien_photonic_2009, pan_multiphoton_2012, aharonovich_solid-state_2016}. While quantum-light sources based on mature solid-state platforms, such as GaAs or diamond, have reached a high level of technological readiness \cite{wei_deterministic_2014, scholl_resonance_2019, wang_boson_2019, tomm_bright_2021, aharonovich_diamond-based_2011, sipahigil_integrated_nodate, riedel_deterministic_2017, janitz_cavity_2020, bersin_telecom_2024, yurgens_cavity-assisted_2024}, single-photon emitters arising in novel 2D materials – especially in transition metal dichalcogenides (TMDs) – have been discovered relatively recently \cite{tonndorf_single-photon_2015, srivastava_optically_2015, he_single_2015, koperski_single_2015, chakraborty_voltage-controlled_2015}. Particularly, these materials have triggered a vast interest in the domain of quantum optics and all-optical information processing due to several factors: the relative affordability of the fabrication compared to III-V technologies \cite{branny_deterministic_2017, drawer_monolayer-based_2023}, the ease of their integration with photonic \cite{iff_purcell-enhanced_2021} and plasmonic structures \cite{luo_deterministic_2018, tugchin_photoluminescence_2023, timmer_ultrafast_2025}, as well as the ability to deterministically seed such emitters via engineered strain \cite{branny_deterministic_2017, paralikis_tailoring_2024} or using electrostatic approaches \cite{thureja_electrically_2024}. 
Moreover, the emission properties of TMD single photon sources can be efficiently tuned by gating \cite{brotons-gisbert_coulomb_2019} and via introducing additional local strain \cite{iff_strain-tunable_2019}. However, advanced applications of TMD–quantum emitters in quantum technologies still remain a significant challenge given the fact that their emission typically suffers from decoherence  \cite{mitryakhin_engineering_2024}. 

While the phenomenon of dominant spectral diffusion is usually attributed to the susceptibility of localized excitons to their environment and especially to uncontrolled charge traps, the high degree of phonon coupling in 2D layers additionally enhances the phonon-induced decoherence \cite{PhysRevB.109.245304, steinhoff_impact_2025}. On the other hand, the detrimental impact of strong Raman sideband emission on the overall coherence could be significantly reduced by using specifically designed photonic structures to mitigate the impact of phonon coupling \cite{mitryakhin_engineering_2024}. It is possible that the radiative relaxation rate of the system through zero-phonon channels can be enhanced by high-quality factor microcavities \cite{grange_reducing_2017}. Furthermore, to counteract environmental effects, and at the same time achieve on-demand performance with minimal emission timing jitter, two-level quantum emitters perform best when being driven resonantly. 

    \begin{figure*}[t]            
        \includegraphics[width=\linewidth]{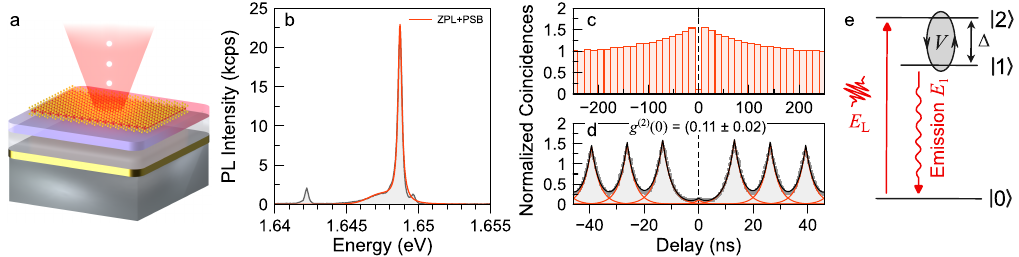}
            \caption{\label{fig:one} (a) Artistic render of the WSe$_2$ ML placed on a Si$_3$N$_4$/SiO$_2$/Au/GaAs slab. The laser beam is focused onto the QD area; the emitted single photons are depicted as white spheres. (b) PL spectrum of an emitter under study. (c) and (d) Second-order correlation of the emission under pulsed excitation. The solid red lines in (d) are fitted double exponential curves, while the black line is sum of these curves. (e) Scheme of the energy level structure of a QD. The emitter is driven via its excited state $\ket{2}$, and PL emission from the exciton ground state $\ket{1}$ is collected. The states $\ket{1}$ and $\ket{2}$ are separated by an energy difference $\Delta$ and coupled via Coulomb interaction with matrix element $V$. 
            }
    \end{figure*}

Resonant excitation of solid-state quantum emitters can be a challenging task due to the necessity to carefully separate a weak signal from the dominant pump laser - either by  polarization filtering \cite{somaschi_near-optimal_2016,ding_-demand_2016, wang_towards_2019, yurgens_cavity-assisted_2024} or by spatial degrees of freedom \cite{makhonin_waveguide_2014}. The requirement for accurate emission-excitation energy matching makes resonant excitation schemes particularly challenging for systems that are prone to spectral diffusion. Fluctuations in the transition energy have been observed even under strictly resonant monochromatic excitation of WSe$_2$-based emitters \cite{kumar_resonant_2016, errando-herranz_resonance_2021}, leading to significant variations in the resonance fluorescence intensity \cite{kumar_resonant_2016}. To our knowledge, coherent, strictly resonant pulsed excitation of TMD quantum dots (QDs) has not yet been demonstrated. % due to experimental complexities. 

On the other hand, there are alternative approaches to resonant excitation that offer greater flexibility when it comes to filtering out excess laser scattering. These include two-photon excitation that is commonly applied for the resonant injection of biexcitons in InGaAs QDs  \cite{jayakumar_deterministic_2013}. %[ D. Huber Natcomm 2016 Highly indistinguishable and strongly entangled photons from symmetric GaAs quantum dots] 
In addition, dichromatic excitation schemes have recently been successfully applied to drive InAs QDs (both symmetrically and asymmetrically with regards to the emitter’s zero phonon line (ZPL)) \cite{he_coherently_2019, karli_super_2022}. The aforementioned methods, in turn, require highly precise engineering of laser pulses, emission-excitation detunings and a high degree of knowledge concerning the emitter coherence and relaxation dynamics. 

In this work, we establish control over thr exciton ground-state population of a quantum emitter in a WSe$_2$ monolayer (ML) by coherently driving it into its excited state. In order to probe the energy level structure of the emitter, photoluminescence excitation (PLE) spectroscopy has been utilized. Consequently, while resonantly pumping into one of the PLE-revealed electronic resonances, we observe an oscillatory behavior in the emission intensity as a function of the pump field amplitude, which is strongly detuning-dependent. Finally, we explain this behavior within the framework of a microscopic model that captures the system dynamics via solving the Lindblad master equation for our effective three-level system. 

The sample under the investigation is a WSe$_2$ ML flake that has been exfoliated from a bulk crystal via the Scotch-Tape method and then transferred via dry-gel stamping \cite{castellanos-gomez_deterministic_2014} onto a planar heterostructure. This structure acts as an effective dielectric antenna, promoting a large portion of the photoluminescence (PL) into the first detection lens, as depicted schematically in Fig.~\ref{fig:one}a. The slab is composed of a GaAs substrate, a 100 nm thick Au layer, a 400 nm thick SiO$_2$ layer and a Si$_3$N$_4$ cap layer with a thickness of 180~nm. During the room-temperature dry transfer, the flake has been deliberately exposed to a large mechanical force to create a plethora of crystalline defects and a stochastic landscape of strain potential. This procedure introduces discrete quantum emitters contrasting the optical response of the free excitons in 2D layers. These emitters have been identified as being of excitonic origin in the presence of tight local confinement \cite{tonndorf_single-photon_2015, koperski_single_2015,he_single_2015, chakraborty_voltage-controlled_2015,srivastava_optically_2015}, and in the following will be referred to as QDs. 

In order to study the emission properties of the QDs in this work, we make use of micro-photoluminescence ($\mu$PL) spectroscopy in confocal geometry, enabling us to probe individual emitters with high spatial resolution determined by a laser profile of 1.5~\textmu m in diameter. Throughout the entire study, the flake is kept in a closed-cycle cryostat at a temperature of 3.8 K. Here, we focus on the emission that is spectrally separated from the ML free exciton or trion emission, i.e. allocated to a spectral range of less than 1.6755~eV. One of such isolated emission lines of a QD with the exciton ground state energy of 1.649~eV is depicted in Fig.~\ref{fig:one}b under conditions, when the QD was driven with a continuous wave (CW) Ti:Sapphire laser of energy $E_{\text{L}}$~=~1.707~eV (detuned from the emission energy $E_1$ by 58~meV) and a power of 1.5~\textmu W. The signal comprises a dominant ZPL with a linewidth of 100 \textmu eV and a broad, underlying acoustic phonon induced feature (phonon sideband, or PSB), which was discussed in detail in  previous works \cite{mitryakhin_engineering_2024, steinhoff_impact_2025, helversen_temperature_2023}. 

The quantum character of the PL from the QD was studied by accessing its second-order correlation function via the Hanbury Brown-Twiss (HBT) experiment. The emitter is driven to the higher-energy state (as will be confirmed later), separated from the exciton ground state by 58~meV, using 2-ps Ti:Sapphire laser pulses at the energy of $E_{\text{L}}$~=~1.707 eV (pulse repetition rate is 76.4~MHz). Here, we use a set of single-photon detectors and a correlation electronics with an overall timing jitter of 250~ps. The resulting second-order correlation function is presented in Fig. \ref{fig:one}(c, d) and features a value at zero delay of 0.11$\pm$0.02, clearly manifesting the single-photon character. Notably, the broad-overview correlation chart further shows a superimposed time-decay constant on the order of 100 ns, which could suggest the involvement of a metastable state in the relaxation dynamics of the investigated system. 

Under these conditions, we consider the emitter to be driven resonantly into the excited exciton state $\ket{2}$ while the emission is detected from the Coulomb-coupled exciton ground state $\ket{1}$ (separated from $\ket{2}$ by an energy difference $\Delta$) according to the energy level structure depicted in Fig. \ref{fig:one}e, which we will evaluate within this study.

In order to scrutinize the internal level structure of the QDs in our sample, we perform PLE spectroscopy. We use a tunable Ti:Sapphire laser in CW mode scanning it from an energy $E_{\text{L}}=1.724$~eV towards the emission energy $E_{\text{1}}$ of our emitters, while monitoring the QD PL. % as a function of the  wavelength. %, which is near the resonance of free exciton in ML $E_{\text{FX}}=1.741$~eV 1.741 eV,

\begin{figure}[t!]
            \includegraphics[width=0.48\textwidth]{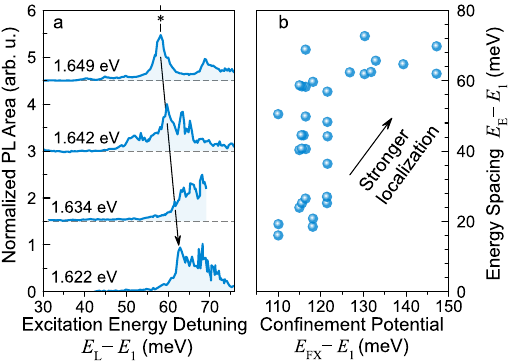}
            \caption{\label{fig:two} (a) PLE spectra of four different QD emitters. The ground-state energies $E_{1}$ are shown beside the respective spectra. The star symbol points to the resonance associated with the excited state of the QD. (b) Statistics of the energy spacing between PLE resonances, attributed to QD higher-energy states, and exciton ground state $E_\text{E}-E_\text{1}$ as a function of effective confinement potential $E_\text{FX}-E_\text{1}$.}
    \end{figure}

The result of this study is shown in Fig. \ref{fig:two}a for four selected emitters.
We observe multiple resonances emerging, most prominently in the 40–50~meV range of laser–ground-state detuning $E_{\text{L}}-E_{\text{1}}$. The strongest resonance was found at an energy of 58 meV above the 1.649 eV QD ground state emission (indicated by the star in the top spectrum in Fig. \ref{fig:two}a). Our spectral analysis indicates that this resonance originates from a higher-lying electronic shell of the QD, as there are no one- or multi-phonon processes in WSe$_2$ that were reported to occur at such an energy. This assignment is further consolidated by analyzing the energy of the excited states $E_{\text{E}}$ (identified in our PLE study) in relation to the energy of the exciton ground state $E_{\text{1}}$. In Fig. \ref{fig:two}b, we therefore plot energy spacing $E_\text{E}-E_\text{1}$ as a function of the effective confinement potential, estimated as the difference between the energy of free exciton in ML $E_{\text{FX}}=1.741$~eV and $E_{\text{1}}$. Indeed, the direct correlation between the localization degree and the energy spacing is present, which establishes the resonances in the PLE measurements as real electronic states.
%emitters, which possess lower emission energies (thus, experience stronger confinement) systematically encounter increasing energy offsets between higher energy of the higher resonances, which establishes these transitions as real electronic resonances. the states as electronic correlation

\begin{figure*}[t]
            \includegraphics[width=1.0\textwidth]{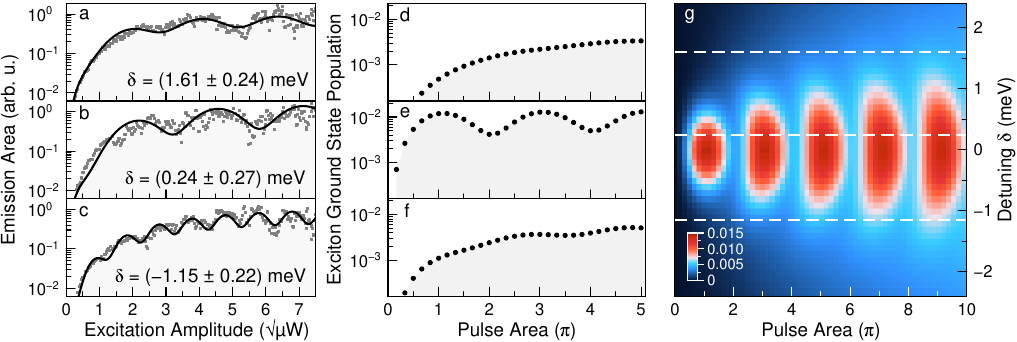}
            \caption{\label{fig:three} Normalized area of PL emission (gray squares) vs square root of laser power measured before microscope objective in cases of different detunings $\delta$ between laser and the excited state of the QD: (a) -1.15 $\pm$ 0.22 meV, (b) 0.24 $\pm$ 0.27 meV, (c) 1.61 $\pm$ 0.24 meV. Solid lines are based on an adapted saturation model. 
            (d-f) display the outcome of a full quantum treatment of the coupled level model for the same detunings. (g) Population of the ground exciton state (color-coded) calculated at specific pulse areas ($x$-axis) and detunings $\delta$ ($y$-axis). The white dashed lines denote the detunings corresponding to the curves shown in (d-f).}
    \end{figure*}

The resonant driving of electronic transitions in discrete two-level or multi-level systems with varying pulse area is expected to give rise to coherent Rabi oscillations, which manifest themselves in the emitted PL, since the emission intensity is proportional to the excited state population. In this study, we record several datasets to investigate the emission of the WSe$_2$ QDs as a function of excitation power, while the system is being driven with picosecond pulses in the spectral proximity of its excited state. In the following, we limit ourselves to one particular emitter with $E_1 = 1.649$~eV, the emission spectrum of which is plotted in Fig. \ref{fig:one}b. The corresponding PLE spectrum with the resonance $E_2=1.707$~eV is shown in the top row of Fig.~\ref{fig:two}a. As expected for a solid-state quantum emitter, the overall population encounters a saturation behavior (see Fig. \ref{fig:three}a-c). This global behavior remains relatively unaffected in our experiment, in which we vary the detuning of the pump laser from the QD excited state $\delta=E_\text{L}-E_2$ in the range of a few meV. However, we encounter an additional effect in our experiments, which is the manifestation of a superimposed oscillatory behavior occurring with great consistency. As we plot in Fig. \ref{fig:three}a-c, the frequency and the amplitude of this oscillation depends significantly on the laser-emitter detuning $\delta$, which hints at a resonant phenomenon that dictates the population dynamics in the QD. In order to highlight the features of the data we plot it alongside with an empirical function $f(x) = A \frac{x^2}{x^2+b}(\sin{(R\cdot{x})}+C)$ (solid lines in \ref{fig:three}a-c). This function accounts for both the saturation and the fundamental oscillatory behavior of the emission intensity. 

 We model the emitter as a three-level system with the basis states $\left\{ \ket{0}, \ket{1},\ket{2} \right\}$ (see Fig.~\ref{fig:one}e), where $\ket{0}$ denotes the system ground state (empty lattice), $\ket{1}$ denotes the exciton ground state and  $\ket{2}$ denotes the excited state of the exciton respectively:
    \begin{equation}
        \begin{split}
        H&=\varepsilon_{1}\sigma_{11}+(\varepsilon_{1}+\Delta)\sigma_{22}+H_{\textrm{pump}}+H_{\textrm{Coul}} \,.
        \end{split}
        \label{eq:Hamiltonian}
    \end{equation}
Here, $\sigma_{ij}=\ket{i}\bra{j}$ and the energy difference between the two exciton states is given by $\Delta$. The emitter is subject to a coherent optical pump described by
    \begin{equation}
        \begin{split}
        H_{\textrm{pump}}=\frac{\hbar f(t)}{2}\sigma_{20}+\frac{\hbar (f(t))^*}{2}\sigma_{02}
        \end{split}
        \label{eq:Hamiltonian_pump}
    \end{equation}
with the time-dependent function
    \begin{equation}
        \begin{split}
        f(t)= \frac{2A}{\sqrt{2\pi\tau^2}}e^{-i\frac{E_{\textrm{L}}}{\hbar} t }e^{-\frac{(t-t_0)^2}{2\tau^2}} \,.
        \end{split}
        \label{eq:pump_function}
    \end{equation}
The pulse area is given by $A$ and the pulse duration is determined by $\tau$. The latter corresponds to the full-width-at-half-maximum of the pump field intensity divided by $2\sqrt{\textrm{ln}2}$. The center $t_0$ of the Gaussian envelope is chosen such that $f(0)=10^{-6}f(t_0)$.
The exciton states are assumed to be coupled via Coulomb configuration interaction, which is described by:
    \begin{equation}
        \begin{split}
        H_{\textrm{Coul}}=V\sigma_{21}+V^*\sigma_{12}\,.
        \end{split}
        \label{eq:Hamiltonian_coul}
    \end{equation}
Note that the configuration interaction leads to mixing of exciton states as well as renormalized energies given by $\varepsilon_{\pm}=\varepsilon_1+\frac{1}{2}\left(\Delta\pm\sqrt{\Delta^2 +4|V|^2} \right) $. We therefore adjust the detuning of the coherent pump with respect to the upper renormalized energy $\varepsilon_+$.

The dynamics of the system is governed by the von Neumann-Lindblad (vNL) equation for the density matrix $\hat{\rho}$ \cite{lindblad_generators_1976, breuer_theory_2002}:
    \begin{equation}
        \begin{split}
        \frac{\partial}{\partial t}\hat{\rho}(t)= -\frac{i}{\hbar}\left[H, \hat{\rho}(t) \right]+\mathcal{L}\hat{\rho}(t)\,.
        \end{split}
        \label{eq:vNL}
    \end{equation}
Lindblad terms describing pure dephasing of the exciton states are given by $\mathcal{L}\hat{\rho}(t)=\big[\mathcal{L}_{2\gamma_1}(\sigma_{11})+\mathcal{L}_{2\gamma_2}(\sigma_{22})\big]\hat{\rho}(t) $
with the general form
$\mathcal{L}_{\gamma}(\hat{O})\hat{\rho}=-\frac{\gamma}{2}\big[
\hat{O}^{\dagger}\hat{O}\hat{\rho}+\hat{\rho}\hat{O}^{\dagger}\hat{O}-2\hat{O}\hat{\rho}\hat{O}^{\dagger}
\big] $.
We solve the vNL equation in a rotating frame by applying a unitary transformation $U(t)=\textrm{exp}(\frac{i}{\hbar}\varepsilon_1 t (\sigma_{11}+\sigma_{22}))$ to the system Hamiltonian (\ref{eq:Hamiltonian}). The temporal width of the pulse and the splitting between the ground and excited states were set in accordance to the experimental values (2 ps FWHM and $\Delta=58$ meV, respectively). The pure dephasing rate $\gamma$ is estimated from the ZPL linewidth as $0.1$ ps$^{-1}$, which is in agreement with the previous study \cite{mitryakhin_engineering_2024}. The Coulomb coupling matrix element $V$ is set to 4 meV as a result of slight adjustment of a typical value to the experiment \cite{PhysRevB.60.5597, Baer2004}. For each pump detuning and pulse area, we propagate the vNL equation until after the pump pulse and extract the final population of the exciton ground state $\ket{1}$.

The outcome of the calculations is plotted in Fig. \ref{fig:three}(d-f), where we present results corresponding to the laser-excited state detuning $\delta$ in the experimental data in Fig. \ref{fig:three}(a-c). From a qualitative point of view, the model captures the main effects which we observe in our experimental study: For a small detuning $\delta \approx 0$ (see Fig. \ref{fig:three}, panels b and e), a clear oscillatory behavior emerges in both experiment and theory. As the detuning acquires finite negative (see Fig. \ref{fig:three}, panels a and d), or positive (see Fig. \ref{fig:three}, panels c and f), values exceeding 1 meV, the oscillation amplitude is damped, and furthermore, the overall intensity in the first maximum of the oscillation is significantly reduced. Both effects can be directly associated with less effective driving of the excited state with increasing detuning.

To clarify this further, in Fig.~\ref{fig:three}g, we plot consistent calculation of the outcome of our model in finer increments of $\delta$, where the clear emergence of oscillating behavior can be recognized for modest detunings. It is important to note, that a direct phonon-assisted transition between the states $\ket{1}$ and $\ket{2}$ can be excluded as a possible explanation, since the phonon modes lack states with energies resonant with the energy spacing $\Delta$, and multiphonon processes are highly inefficient. %It is important to note, that our model a priori rules out a direct phonon transition between the exciton excited state  $\ket{2}$ and the exciton ground state  $\ket{1}$, since for a reasonable choice of phonon energies and dephasing parameters, no oscillatory behavior can be reproduced.
Finally, we point out that while we do observe a significant dependence of the oscillation period on $\delta$ in our experiments, we cannot incorporate this effect into our model.  While it remains somewhat speculative at the present stage, we believe that this effective renormalization of the pulse area may be captured via a full microscopic treatment of all phonon modes in our system.  

%\section{Conclusion}

In conclusion, we demonstrate the emergence of Rabi oscillations in a single TMD quantum emitter, which is resonantly excited in a higher electronic shell. The characteristic oscillatory behavior is found to depend on the effective laser detuning. %  in conjunction with phonon dephasing processes, which are fully captured in our microscopic quantum optical model. 
While our work clearly shows the great potential of TMD quantum emitters arising from their strong light-matter coupling,  further efforts towards coherent driving of the electronic ground state transition, and possibly combining coherent excitation methods with high-quality optical cavities should be undertaken to fully harness their strength. 

\section{Acknowledgements}

Support by the German Research Foundation (DFG) within the program INST 184/234-1 and the priority program SPP2244 via the projects Gi1121/4-2 is acknowledged. The work was funded by Federal Ministry of Research, Technology and Space (BMFTR) within the projects EQUAISE (Project No 13N16354), TubLan Q.0 (Project No 16KISQ088), and Comphort (Project No 16KIS2107). C. S. acknowledges the support by DFG (Grant No. INST184/220-1 FUGG). J.L. was supported by the Knut \& Alice Wallenberg Foundation through the Wallenberg Centre for Quantum Technology (WACQT). A.P. would like to acknowledge the Swedish Research Council (grant 2021-04494) and Carl Tryggers Stiftelse (grant CTS 24:3526).
This work was supported by the QuantERA II Program that
has received funding from the EU Horizon 2020 research and
innovation program under GA No 101017733.

\bibliography{2025_Rabi}

\end{document}